# Energy efficient and fast reversal of a fixed skyrmions with spin current assisted by voltage controlled magnetic anisotropy

Dhritiman Bhattacharya[1], Md. Mamun Al-Rashid[1,2], Jayasimha Atulasimha[1,2,*]

1. Department of Mechanical and Nuclear Engineering, Virginia Commonwealth University.
2. Department of Electrical and Computer Engineering, Virginia Commonwealth University.

*jatulasimha@vcu.edu

**Abstract:** Recent work [1,2] suggests that ferromagnetic reversal with spin transfer torque (STT) requires more current in a system in the presence of DMI than switching a typical ferromagnet of the same dimensions and perpendicular magnetic anisotropy (PMA). However, DMI promotes stabilization of skyrmions and we report that when the perpendicular anisotropy is modulated (reduced) for both the skyrmion and ferromagnet, it takes much smaller current to reverse the fixed skyrmion than to reverse the ferromagnet in the same time, or the skyrmion reverses much faster than the ferromagnet at similar levels of current. We show with rigorous micromagnetic simulations that the skyrmion switching proceeds along a different path at very low PMA which results in a significant reduction in the spin current required or time required for reversal. This can have potential for memory application where a relatively simple modification of the standard STT-RAM to include a heavy metal adjacent to the soft magnetic layer and with appropriate design of the tunnel barrier can lead to energy efficient and fast magnetic memory device based on the reversal of fixed skyrmions.

1. **Introduction:**

Dzyaloshinskii-Moriya Interaction [3, 4] (DMI) present at a ferromagnetic thin film and heavy metal (with large spin orbit coupling) interface can arrange the spins in a topologically protected spiral orientation known as magnetic skyrmions. Such a state was predicted in magnetic thin films and multilayers [5] and experimentally realized a decade later [6-12]. Recent reports also include demonstration of room temperature stabilization of skyrmion lattice [9], and isolated skyrmions [19]. Magnetic skyrmions can be moved with a small current [13] and hence racetrack memory and logic gates based on current induced skyrmion motion were proposed and extensively studied [14-16]. Furthermore, demonstration of creation and annihilation of skyrmions in nanostripes using spin current [17] and electrical voltage [18] have also been shown experimentally. However, such devices based on moving skyrmions could have a large foot print. To overcome this limitation, we propose a method that utilizes core reversal of a fixed skyrmion confined in a nanostructure, which could result in high density memory elements. A number of techniques have been proposed to achieve core reversal of magnetic skyrmions e.g. using a magnetic field [20], microwaves [21], electrical voltage [22, 23], and spin current [24]. $I^2R$ loss in spin current induced magnetization reversal, however, can be considerably large. Reducing this loss requires reduction of the switching current, which can be achieved by employing methods to reduce perpendicular magnetic anisotropy (PMA), which temporarily depresses the energy barrier between the "up" and "down" state during spin current induced switching. This can improve the energy-efficiency and switching speed without compromising the thermal stability of the nanomagnet based computing device. In this paper, we propose such a hybrid scheme where application of a small voltage can reduce the threshold current needed for reversal of skyrmions, present rigorous micromagnetic simulation to explain the underlying physics of the reversal process, and quantify improvements in terms of dissipated energy and switching speed.

One of the ways of reducing magnetic anisotropy using a voltage is VCMA (Voltage Control of Magnetic Anisotropy), where an external voltage modulates the electron density at a ferromagnet/oxide interface and thereby causing a change in the PMA [25, 26]. The subsequent reduction in switching current utilizing VCMA has been shown for magnetization reversal of ferromagnets with uniform magnetization orientation [27, 28]. In this paper, we examine the viability and investigate the physics of reversal of the magnetization state of a fixed skyrmion using such a hybrid scheme. The presence of DMI in a skyrmion system is what distinguishes our current study from those performed on uniformly magnetized ferromagnetic systems.

A simplified Magnetic Tunnel Junction (MTJ) structure for the implementation of our proposed hybrid scheme for switching a magnetic skyrmion (free layer of the MTJ) is shown in Fig. 1, where the fixed layer is a perpendicularly magnetized ferromagnet. One common pair of electrodes is used for application of both VCMA and spin current. Such Heavy Metal/Ferromagnet/Insulator trilayer can have epitaxial strain, which has been shown to give rise to a ∧-shaped electric field dependence of magnetic anisotropy [29, 30]. As a result, application of a voltage regardless of the polarity will reduce the PMA. However, the direction of the spin current will depend on the polarity of the applied voltage. Skyrmions can be reversed with a spin current when $\vec{m}_p \cdot \vec{m}_c = 1$, where $\vec{m}_p$ is the polarity of spin current and $\vec{m}_c$ is the polarity of the skyrmion core [24]. This is very convenient since we can change the polarity of the applied voltage depending on the required direction of the spin polarized current for reversal and nevertheless achieve PMA reduction due to VCMA. Therefore, if we use a fixed layer with magnetization pointing up (down), we can reverse a skyrmion with core pointing up applying a positive (negative) voltage. Hence, this basic MTJ structure is sufficient to carry out our switching scheme given that an epitaxial strain and therefore a ∧-shaped electric field dependence of magnetic anisotropy is present in the structure. This is well suited to typical magnetic memory applications as this the structure is similar to existing spin transfer torque (STT) RAM and only requires appropriate design of the MgO layer thickness and addition of a heavy metal layer between the free layer and the substrate. Reversing the skyrmion will change the magnetoresistance of the ferromagnet (hard layer)/tunnel barrier/skyrmion (soft later) MTJ structure appreciably, thereby allowing the skyrmion state to be read easily (more detail in supplement).

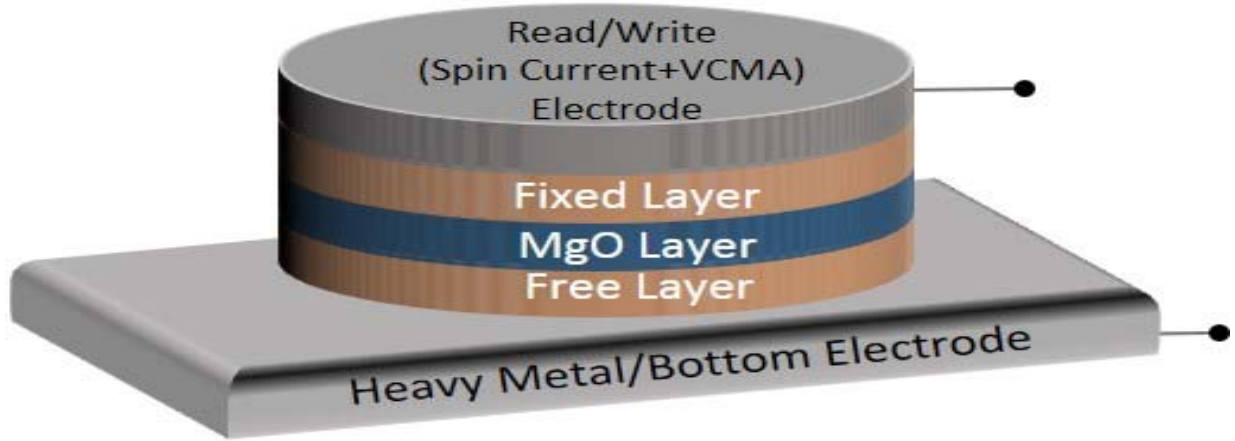

**Figure 1.** Simplified MTJ structure

## 2. Micromagnetic Simulation:

Micromagnetic simulation software-OOMMF was used to perform the simulations [31] where the magnetization dynamics is simulated using the Landau-Lifshitz-Gilbert (LLG) equation:

$$\frac{\partial \vec{m}}{\partial t} = \vec{\tau} = \left(\frac{-\gamma}{1+\alpha^2}\right)\left(\vec{m} \times \vec{H}_{eff} + \alpha\left(\vec{m} \times (\vec{m} \times \vec{H}_{eff})\right)\right) \quad (1)$$

where $\vec{m}$ is the reduced magnetization ($\vec{M}/M_{sat}$), $M_{sat}$ is the saturation magnetization, $\gamma$ is the gyromagnetic ratio and $\alpha$ is the Gilbert damping coefficient. The quantity $H_{eff}$ is the effective magnetic field which is given by,

$$\vec{H}_{eff} = \vec{H}_{demag} + \vec{H}_{exchange} + \vec{H}_{anis} \quad (2)$$

Here, $H_{demag}$ is the effective field due to demagnetization energy, $H_{exchange}$ is the effective field due to Heisenberg exchange coupling and DMI interaction.

The DMI contribution to the effective exchange field is given by [31]:

$$H_{DM} = \frac{2D}{\mu_0 M_{sat}}[(\vec{\nabla} \cdot \vec{m})\hat{z} - \vec{\nabla} m_z] \quad (3)$$

where $m_z$ is the z-component of magnetization and D is the effective DMI constant.

$H_{anis}$ is the effective field due to the perpendicular anisotropy evaluated in the OOMMF framework in the manner described in Ref 32.

$$\vec{H}_{anis} = \frac{2K_{u1}}{\mu_0 M_{sat}}(\vec{u} \cdot \vec{m})\vec{u} + \frac{4K_{u2}}{\mu_0 M_{sat}}(\vec{u} \cdot \vec{m})^3 \vec{u} \quad (4)$$

where, $K_{u1}$ and $K_{u2}$ are first and second order uniaxial anisotropy constants and $\vec{u}$ is the unit vector in the direction of the anisotropy (i.e. perpendicular anisotropy in this case).

VCMA effectively modulates the anisotropy energy density, which is given by $\Delta$PMA = $a$E. Here $a$ and E are respectively the coefficient of electric field control of magnetic anisotropy and the applied electric field. The resultant change in uniaxial anisotropy due to VCMA is incorporated by modulating $K_{u1}$ while keeping $K_{u2} = 0$.

The torque due to spin current is given by,

$$\tau_{STT} = \gamma \beta \left(\vec{m} \times (\vec{m} \times \vec{m}_p)\right) \quad (5)$$

$$\beta = \frac{h}{2\pi} \frac{PJ}{2\mu_0 e M_{sat} L} \quad (6)$$

Here, $\vec{m}_p$ is the unit vector of spin polarization direction, $h$ is Planck's constant, P is the degree of spin polarization, J is the current density, $\mu_0$ is vacuum permeability, e is the electron charge, L is the thickness of the free layer. In this study, current is assumed to be uniform along the diameter of the nanodisk. For the sake of simplicity, field like torque and Oersted field due to current flow is not included. However, as these terms are consistently not considered for all the cases simulated, the key conclusions of this study will not change significantly even if these terms are considered.

If the pinned ferromagnetic layer has a high enough PMA, the spin current or the VCMA will not be able to rotate its magnetization significantly. Therefore, modeling the magnetization dynamics in the pinned layer can be avoided without compromising the accuracy of the simulation (see supplementary section for further discussion). Hence, we only simulated the free layer of an MTJ, which was chosen to be a nanodisk of 100 nm diameter and 0.8 nm thickness. Our geometry was discretized into 2×2×0.8 nm³ cells. We used typical material parameters of CoFe for the free layer in our simulations listed in the table below compiled from Ref 33 and Ref 34. We used a fixed ferromagnetic layer with magnetization pointing up. All simulations were carried out at T=0 K, i.e. effect of thermal noise on magnetization dynamics was not included in this study. However, we show that skyrmion is stable in the presence of room temperature thermal noise in the supplementary section. (Fig. S3)

TABLE 1. List of Parameters (33, 34).

| Parameters | Value |
|---|---|
| Saturation Magnetization ($M_s$) | $1.3 \times 10^6$ A/m |
| Exchange Constant (A) | $1 \times 10^{-11}$ J/m |
| Perpendicular Anisotropy Constant ($K_{u1}$) | 1.1 MJ/m³ |
| Effective Perpendicular Anisotropy ($K_{eff} = K_{u1} - \frac{\mu_0}{2} M_{sat}^2$) | 43570 J/m³ |
| Gilbert Damping ($\alpha$) | 0.01 |
| DMI Parameter (D) | 1.4 mJ/m² |
| Degree of Spin Polarization (P) | 0.4 |

3. **Results: Magnetization switching in various cases**

For these material parameters, ferromagnetic state is the ground state. Skyrmion can be formed by applying a current pulse and emerges as a stable state once formed [17]. We start with a skyrmionic state with spins in the core pointing upwards and spins in the periphery pointing downwards. As our fixed layer points upwards, a positive current will initiate reversal (as, $\vec{m}_p \cdot \vec{m}_c = 1$). We inject a current pulse of 0.1 ns rise and 0.1 ns fall time (shown in Fig. 2) and find the critical switching current and time required for reversal. Figure 2 shows snapshot of magnetization dynamics during the switching process while a spin current of $1 \times 10^{11}$ A/m² was employed. The spin current excites breathing mode of increasing amplitude. Also, the skyrmion texture continually alters between Néel (radial outward and inward) and Bloch (counter clockwise and clockwise) states [35, 36]. These two motions are synchronized. Thus, the breathing mode stabilizes the Néel skyrmion texture at the largest and the smallest core size and Bloch texture in between these Néel states. This transformation can be seen in the supplemental video (S1_video). Due to this spin wave excitation, the skyrmion core expands and shrinks (Fig. 2, t=3.03 ns, 4.57 ns, 4.87 ns, 5.18 ns, 5.95 ns) and eventually reverses. Once reversal occurs (t=7.18 ns) the torque induced by the spin current acts as a damping agent and skyrmion with opposite core polarity is stabilized (Fig. 2, Final). The critical switching current density is found to be $6.5 \times 10^{10}$ A/m$^2$ and takes 13.6 ns to complete the switching. The values of switching time and critical current are in line with the values reported in Ref 24.

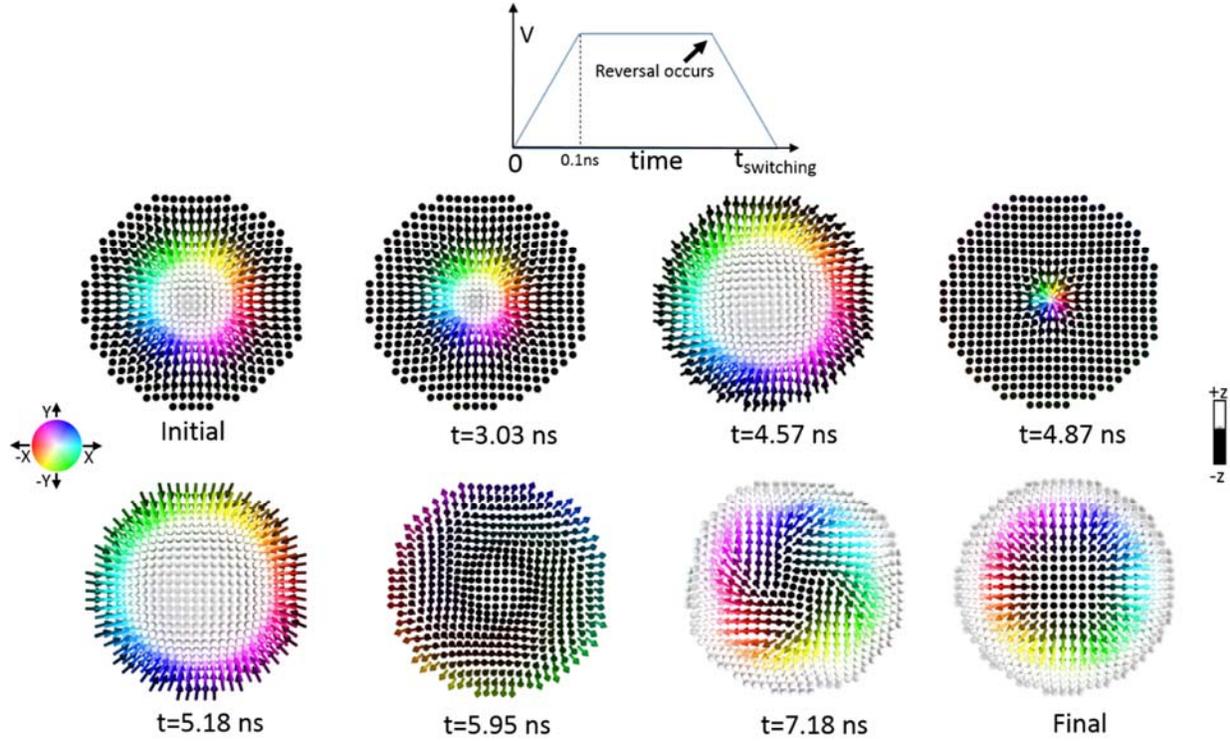

**Figure 2**. Snapshot of magnetization dynamics during the switching process with only spin current ($1 \times 10^{11}$ A/m$^2$). The voltage pulse is shown on top. Note that, the rotational motion of domain wall spins is not shown here for simplicity. [See supplemental video, S1_video for more details].

Next, we study the effect of reducing PMA on the switching behavior. Reduction in PMA creates alternative path for reversal. We studied two cases, 10% and 20% reduction in PMA (which we denote as ΔPMA=10% and ΔPMA=20%). This reduction in perpendicular anisotropy shifts the easy orientation from perpendicular to in plane with $K_{eff} = -66430$ J/$m^3$ for ΔPMA=10% and $K_{eff} = -176430$ J/$m^3$ for ΔPMA=20%. We assumed 0.1 ns rise and 0.1 ns fall time for both the pulses (i.e. spin current and perpendicular anisotropy modulation). Fig. 3 corresponds to the case where ΔPMA=10%. This switching resembles the previous case where only spin current was used to reverse the skyrmion. The reversal stabilizes a skyrmion with opposite polarity, and after restoring the PMA a skyrmionic state exactly opposite to the initial state stabilizes (Fig. 3, Final). Although the switching behavior remains same, the critical current density is reduced by ~1.6 times compared to the case where no VCMA is applied. The switching time vs. current density is discussed later in this paper.

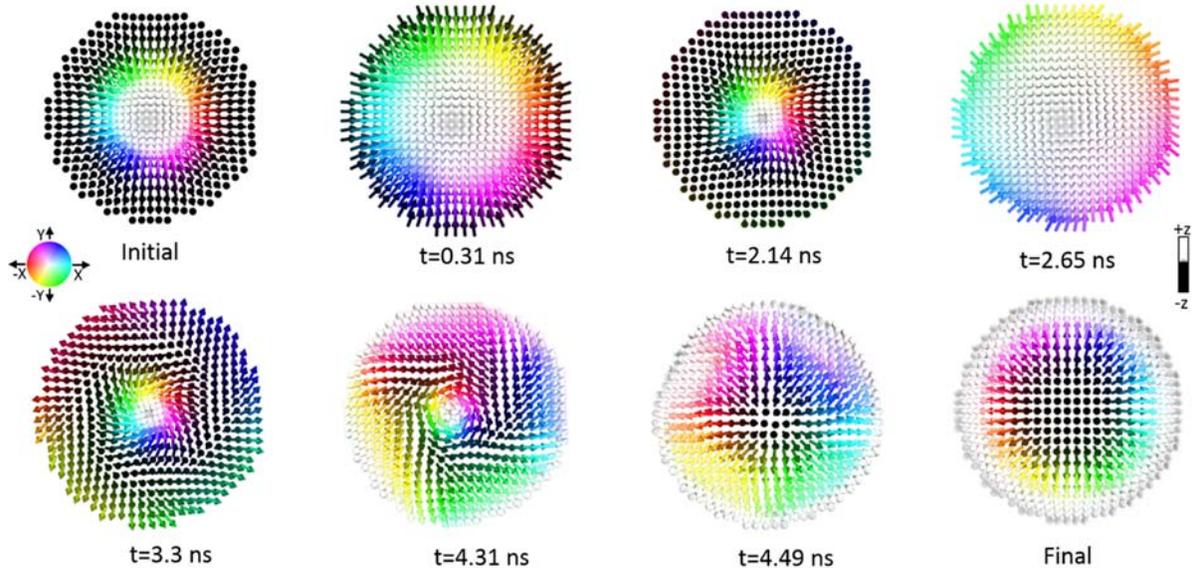

**Figure 3**. Snapshot of magnetization dynamics during the switching process with spin current ($6\times 10^{10}$ A/m$^2$) and small $\Delta$PMA (10%). [See supplemental video, S2_video for more details].

Interestingly, when PMA is reduced further ($\Delta$PMA=20%), the switching follows a very different trajectory which is shown in Fig. 4. At first, reduction in PMA pushes the peripheral spins to the x-y plane while the core still point upwards. (Fig. 4, t=0.1 ns). The spin wave excitation alters the magnetic texture between circular (clockwise: Fig. 4, t=0.7 ns and counter clockwise: Fig. 4, t=1.09 ns) and radial vortex (Fig. 4, t=0.9 ns) states [37] and ultimately reverses the core (Fig. 4, t=1.39 ns). Therefore, a radial vortex state with core pointing downward is formed (Fig. 4, t=1.39 ns). After reversal, restoring PMA pushes the peripheral spin upwards and thus a skyrmion with polarity opposite to the initial state is stabilized (Fig. 4, Final). Here, the critical current density is reduced by ~4.6 times compared to the case where no VCMA is applied along with a drastic reduction in switching time.

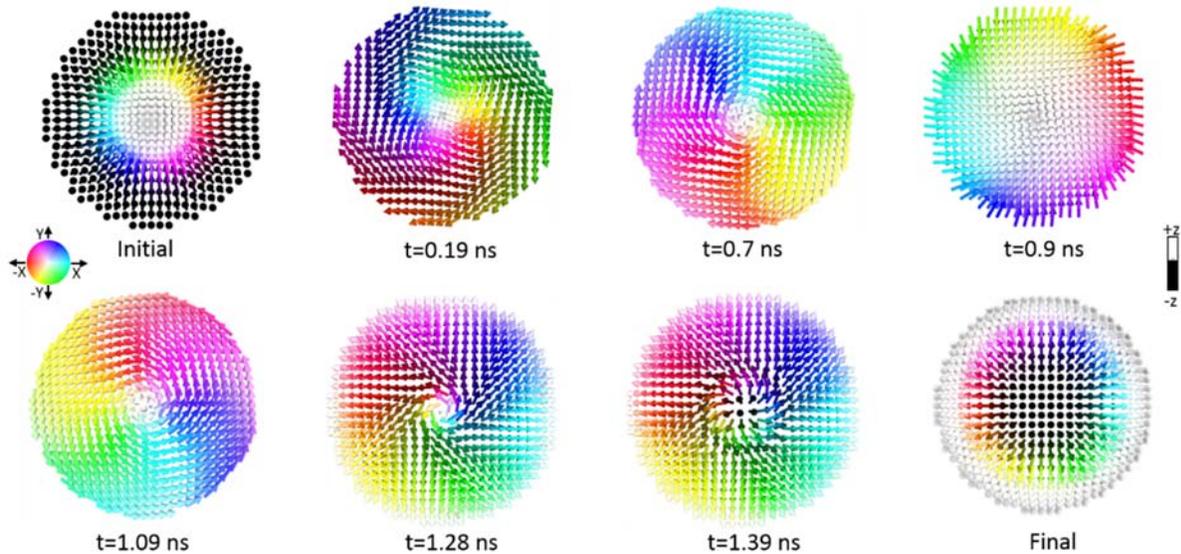

**Figure 4.** Snapshot of magnetization dynamics during the switching process with spin current ($1.4\times 10^{10}$ A/m$^2$) and large $\Delta$PMA (20%). [See supplemental video, S3_video for more details].

Fig. 5 shows switching time vs. current density for the three cases discussed. The critical current and the switching time needed for reversal of the fixed skyrmion is substantially reduced in the latter two cases. Also, with similar level of current, reduction in PMA results in faster switching. We compare this reversal with ferromagnetic reversal in a system with same PMA but no DMI. When, only spin current (i.e. no VCMA is considered) induces reversal, the skyrmionic reversal takes longer time (Fig. 5) than for the ferromagnet without DMI. Furthermore, critical current density is almost 4 times smaller for the ferromagnetic case (not shown in Fig. 5).

However, the skyrmionic reversal shows an improved performance in the hybrid scheme. The skyrmion switching, in the case with highest VCMA ($\Delta$PMA=20%), can take place approximately *five times faster* (1.5 ns vs. 7.7 ns) than the ferromagnetic reversal for current densities of $1.4\times10^{10}$ A/m$^2$. For a fixed switching time ~1.5 ns the current density required to switch the skyrmion is *more than 10 times smaller* than that required to switch the ferromagnet with the same VCMA ($\Delta$PMA=20%) (not shown in Fig. 5). The concomitant write energy (I$^2$R loss) would therefore show *two orders of magnitude improvement*. Hence, one could *write five times as faster* at the same current density or write with *two orders of magnitude less energy* for the same switching time.

Furthermore, if you consider conventional spin transfer torque (STT) devices without VCMA or DMI, then the current density to switch in ~ 1.5 ns is $3\times10^{11}$ A/m$^2$ (not shown in Fig. 5) while the corresponding current density to switch in 1.5 ns for the skyrmion with VCMA is ~$1.4\times10^{10}$ A/m$^2$. Thus, the best case reduction in current density for switching in ~ 1.5ns is about 21 times which can result in ~441 times less energy dissipation.

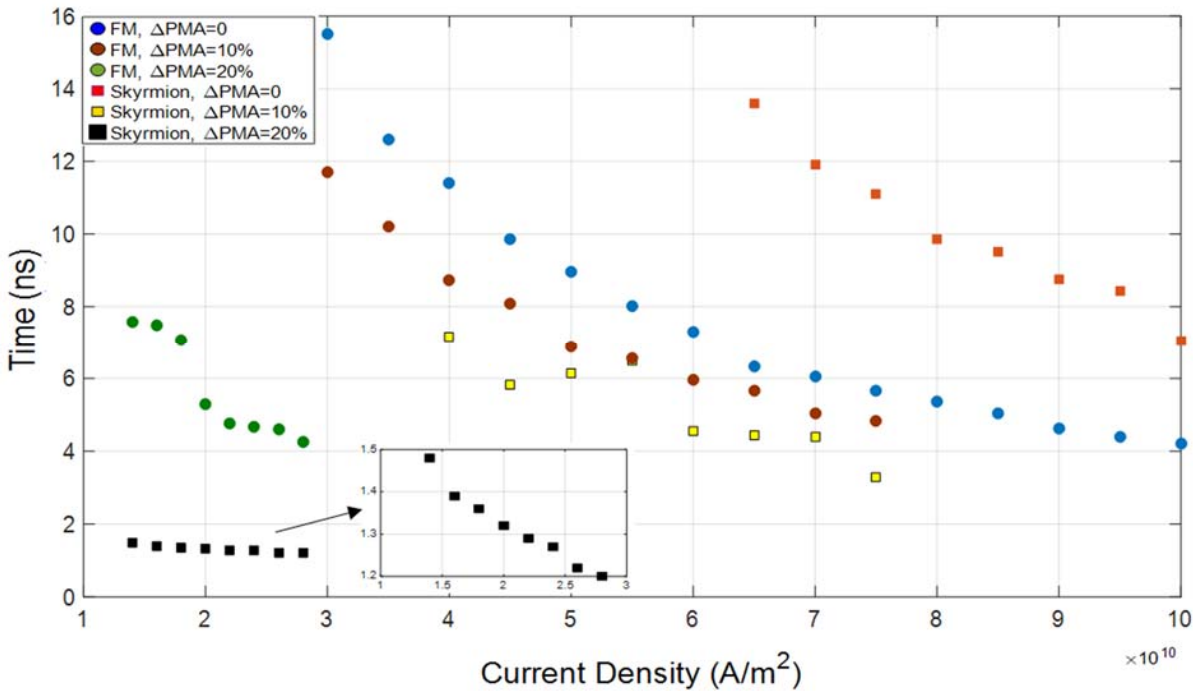

Fig 5. Switching time vs current density for ferromagnetic and skyrmion reversal

## 4. Conclusion

The modulation of the interface anisotropy energy is given by $\Delta PMA = aE$, where $a$ and E are respectively the coefficient of electric field control of magnetic anisotropy and the applied electric field. The coefficient of electric field control of magnetic anisotropy is defined as, $a = \frac{\Delta PMA \times t_{CoFe}}{\Delta V/t_{MgO}}$. The theoretical reported value of "$a$" is 250 $\mu J/m^2$ per V/nm [29]. Thus, with a 0.8 nm thick free layer and 1 nm thick MgO layer, 20% change in the perpendicular anisotropy can be obtained by applying 0.7 volt. The energy required to charge the capacitive MgO layer (relative permittivity ≈ 7 [38], thickness ≈ 1 nm, diameter ≈ 100 nm) is 0.12 fJ which is negligible compared to the typical write energy is conventional spin transfer torque (STT) devices. Thus, the use of VCMA in conjunction with spin current to switch fixed skyrmion based memory devices could result in an order of magnitude smaller energy dissipation compared to switching conventional STT devices or voltage assisted reversal of ferromagnets.

In conclusion, we showed voltage assisted reversal process of skyrmionic state can significantly reduce the write energy over voltage assisted reversal of ferromagnets. Furthermore, comparing this with ferromagnetic switching, we found skyrmion switching induced by spin current can be faster while assisted by a small voltage induced change in perpendicular magnetic anisotropy. Moreover, our proposed device structure can be fabricated with very small modification to the existing spin transfer torque (STT) RAM device. Hence, this work can contribute significantly towards implementing energy efficient non-volatile nanomagnetic memory devices based on existing spin transfer torque (STT) writing schemes.

**References:**

1. P-H Jang, K Song, S-J Lee, S-W Lee, and K-J Lee 2015 Appl. Phys. Lett. 107, 202401.

**2.** J. Sampaio, A. V. Khvalkovskiy, M. Kuteifan, M. Cubukcu, D. Apalkov, V. Lomakin, V. Cros and N. Reyren 2016 Appl. Phys. Lett. 108, 112403

3. I. Dzyaloshinskii, 1958 J. Phys. Chem. Solids 4, 241–255 (1958).

3. T. Moriya 1960 Phys. Rev. 120, 91–98.

3. A. Bogdanov, and U. Rößler, 2001 Phys. Rev. Lett. 87, 037203.

6. S. Mühlbauer, B. Binz, F. Jonietz, C. Pfleiderer, A. Rosch, A. Neubauer, R. Georgii, P. Böni, 2009 Science 323, 915.

7. S. Seki, J. -H. Kim, D. S. Inosov, R. Georgii, B. Keimer, S. Ishiwata, Y. Tokura. 2012 Phys. Rev. B 85, 220406.

8. S. Seki, X. Z. Yu, S. Ishiwata & Y. Tokura, 2012 Science 336, 198–201.

9. S. Woo et al, 2016 Nature Materials 15, 501.

10. S. Heinze, K. V. Bergmann, M. Menzel, J. Brede, A. Kubetzka, R. Wiesendanger, G. Bihlmayer & S. Blügel, 2011 Nat. Phys. 7, 713.

11. X. Z. Yu, Y. Onose, N. Kanazawa, J. H. Park, J. H. Han, Y. Matsui, N. Nagaosa & Y. Tokura, 2010 Nature (London) 465, 901.

12. X. Z. Yu, N. Kanazawa, Y. Onose, K. Kimoto, W. Z. Zhang, S. Ishiwata, Y. Matsui & Y. Tokura, 2011 Nat. Mater. 10, 106.

13. T. Schulz, R. Ritz, A. Bauer, M. Halder, M. Wagner, C. Franz, C. Pfleiderer, K. Everschor, M. Garst & A. Rosch 2012 Nature Phys. 8, 301–304.

14. A. Fert, V. Cros, & J. Sampaio, 2013, Nat. Nanotechnol. 8, 152–156.

15. R. Tomasello, E. Martinez, R. Zivieri, L. Torres, M. Carpentieri & G. Finocchio, 2014 Sci Rep 4, 6784.

16. X.Zhang, M Izawa and Y Zhou, 2015 Sci Rep 5, Article number: 9400 (2015)

17. N. Romming, C. Hanneken, M. Menzel, J. E. Bickel, B. Wolter, K. V. Bergmann, A. Kubetzka, R. Wiesendanger, 2013 Science 341, 636.

18. P-J Hsu, A. Kubetzka, A. Finco, N. Romming, K.V. Bergmann, R. Wiesendanger, 2016 e-print arXiv: 1601.02935.

19. C Moreau-Luchaire, 2016 Nat. Nanotechnol. **11**,444–448.

20. B. Zhang, W. Wang, M. Beg, H. Fangohr and W. Kuch, 2015 Appl. Phys. Lett. 106, 102401 (2015).

21. M. Beg et al., 2015, Sci. Rep. 5, 17137.

22. D. Bhattacharya, M. Al-Rashid, J. Atulasimha, 2016 s*ci. Rep.***6**, 31272

23. Y. Nakatani, M. Hayashi, S. Kanai, S. Fukumi and H. Ohno, 2016 Appl. Phys. Lett. 108, 152403.

24. Y. Liu, H. Du, M. Jia and A. Du, 2015, Phy Rev B 91, 094425.

25. M. K. Niranjan, C.-G. Duan, S. S. Jaswal, and E. Y. Tsymbal 2010 Appl. Phys. Lett. 96, 222504 (2010).

26. P. Amiri, K. Wang, 2012 World Scientific 1240002.

27. Y. Shiota, T. Nozaki, F. Bonell, S. Murakami, T. Shinjo and Y Suzuki 2012 Nat. Mater. 11, 39.

28. W. Wang, M. Li, S. Hageman & C. L. Chien 2012 Nat. Mater. 11, 64.

29. P. V. Ong, N Kioussis, D. Odkhuu, P. K Amiri, K. L. Wang, and G. P. Carman, 2015 Phys. Rev. B 92 020407.

30. Y. Hibino, T. Koyama, A. Obinata, T. Hirai, S. Ota, K. Miwa, S. Ono, F. Matsukura, H. Ohno and D. Chiba 2016 Appl. Phys. Lett. **109**, 082403.

31. S. Rohart and A. Thiaville, 2013, Phys Rev B 88, 184422.

32. M. J. Donahue and D. G. Porter, ``OOMMF User's Guide, Version 1.0,'' NISTIR 6376, National Institute of Standards and Technology, Gaithersburg, MD (Sept 1999).

33. M. Belmeguenai, M. S. Gabor, Y. Roussigne. Stashkevich, S. M. Cherif, F Zighem and C. Tiusan, 2016 Phys Rev B **93**, 174407.

34. S. Emori, U. Bauer, S.-M. Ahn, E. Martinez, and G. S. Beach, 2013, Nat. Mater. 12, 611.

35. M Carpentieri, R Tomasello, R Zivieri & G Finocchio, 2015 Sci Rep 5, 16184.

36. R. H. Liu, W. L. Lim, and S. Urazhdin, 2015 Phys Rev Lett 114, 13720.

37. G. Siracusano, R. Tomasello A. Giordano, V. Puliafito, B. Azzerboni, O. Ozatay, M. Carpentieri, G. Finocchio, 2016 Phys Rev Lett 117, 087204.

38. I. Ho, X. Yu, J. Mackenzie, 1997 Journal of Sol-Gel Science and Technology 9, 295-301

# Supplementary Information

## Energy efficient and fast reversal of a fixed skyrmions with spin current assisted by voltage controlled magnetic anisotropy

Dhritiman Bhattacharya[1], Md. Mamun Al-Rashid[1,2], Jayasimha Atulasimha[1,2,*]

1. Department of Mechanical and Nuclear Engineering, Virginia Commonwealth University.
2. Department of Electrical and Computer Engineering, Virginia Commonwealth University.
*jatulasimha@vcu.edu

**Calculating the change in magnetoresistance**:

Total resistance of the MTJ structure can be written as [1],

$$R = R_0 + \Delta R \, sin^2 \frac{\theta}{2}$$

Here, $\Delta R$ is the increment in magnetoresistance and $\theta$ is the angle between fixed and the free layer. When a free layer is switched between two ferromagnetic states (up and down), $\theta$ changes from 0° to 180° and R varies from $R_0$ to $R_0 + \Delta R$.

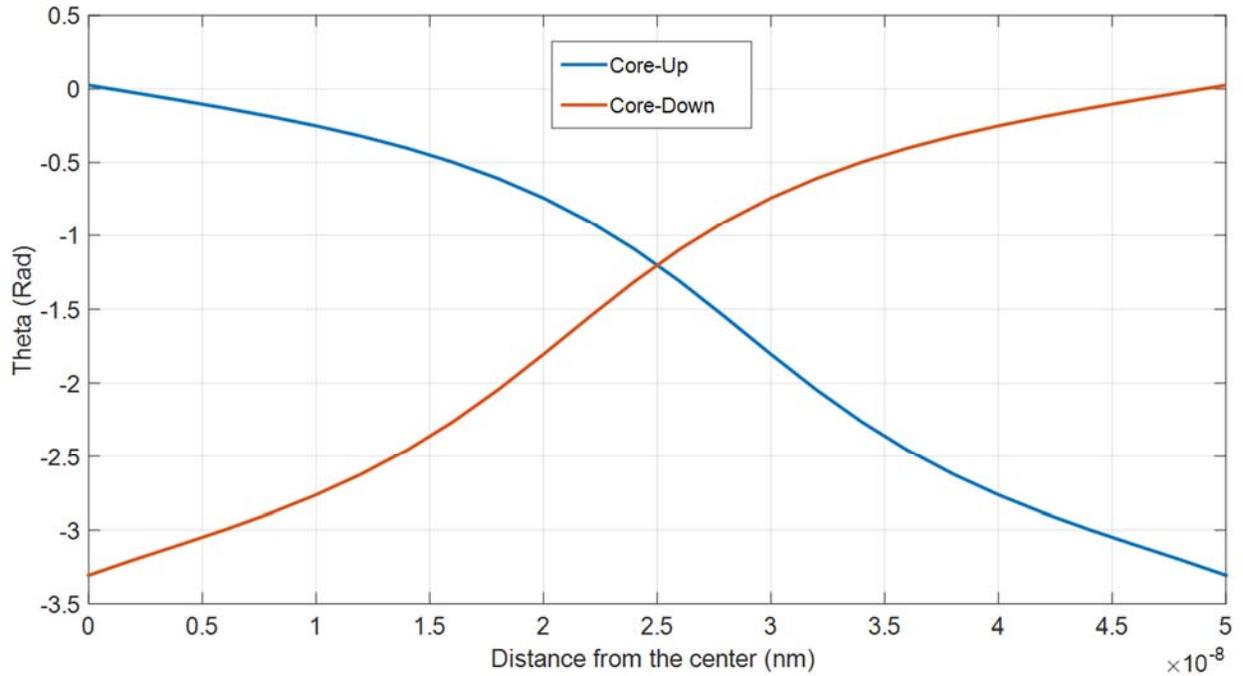

Fig. S1: Theta vs distance from center for two skyrmionic states

We calculated theta along the radius for two skyrmionic states and found R varies from $R_0 + 0.28\Delta R$ to $R_0 + 0.7\Delta R$ between core-down and core-up skyrmions. The magnetoresistance change can be more than 40% compared to the change in ferromagnetic reversal. Hence, the readout signal will change clearly as the magnetization flip from one skyrmionic state to another which should be enough to carry out a successful readout operation.

**Fixed layer response to the current pulse:**

We provide a simulation [Fig. S2] that shows the response of the pinned layer (PMA=2 MJ/m^3) while a current of density $5\times10^{10}$ A/m$^2$ is passed through it for a time period of 10 ns (more than 5 times longer than used in simulations for the switching of the free layer) along with the maximum PMA reduction applied to the free layer. Even then there is no noticeable changes in magnetization of the hard layer.

It is also feasible to experimentally fabricate such high perpendicular anisotropy layers using proper interfacial layers, for example using a Bi layer can give rise to PMA up to 6 MJ/ m$^3$ [2].

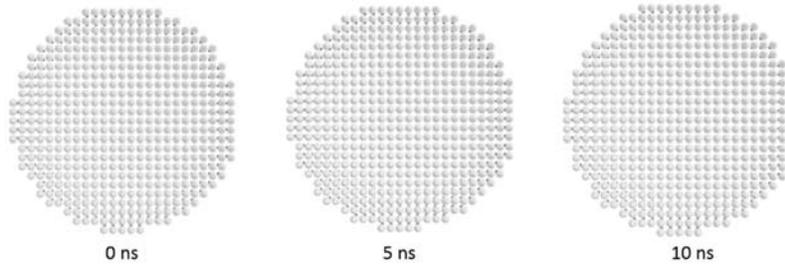

Fig. S2: Response of the pinned layer (PMA=2 MJ/m^3) while a current of density $5\times10^{10}$ A/m$^2$ is passed through it for a time period of 10 ns

**Thermal Stability:**

Thermal stability is indeed a key element in the successful operation of any device. Our simulation (10 snapshots taken at a time interval of 1 ns are shown) showing a stable skyrmion state in the presence of room temperature thermal fluctuation is shown below. Though the skyrmion core fluctuates (moves a bit in space) due to thermal noise, it does not flip spontaneously to an opposite polarity skyrmionic state or a ferromagnetic state.

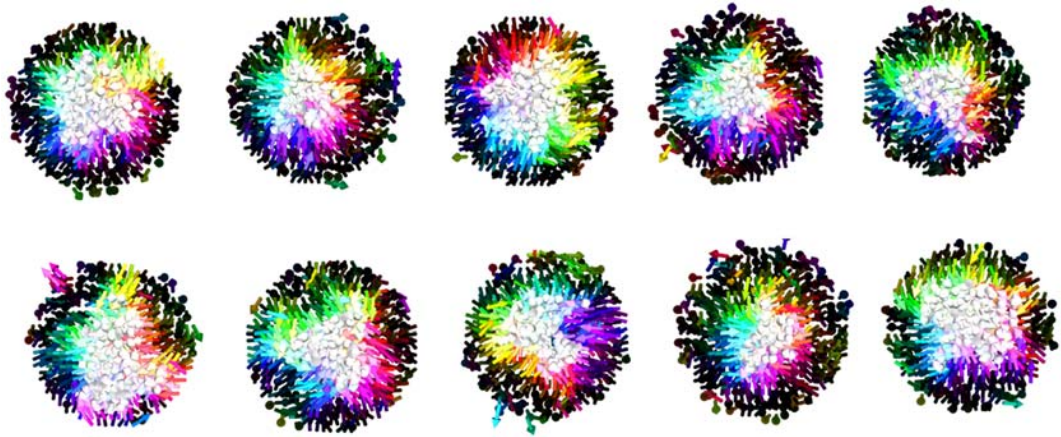

Fig. S3: Snapshots taken at a time interval of 1 ns in the presence of thermal noise

References:

1. Steren LB, Barthelemy A, Duvail JL, Fert A, Morel R, Petroff F, Holody P, Loloee R, Schroeder PA. Phys. Rev. B. 51(1):292 (1995)

2. Peng S, Zhao W, Qiao J, Su L, Zhou J, Yang H, Zhang Q, Zhang Y, Grezes C, Amiri PK, Wang KL, App. Phys. Lett. 110(7):072403 (2017).